\documentclass[cits]{PoS}

\usepackage[mathscr]{euscript}

\newcommand{\BE}{\begin{displaymath}}
\newcommand{\EE}{\end{displaymath}}
\newcommand{\BNE}{\begin{equation}}
\newcommand{\ENE}{\end{equation}}
\newcommand{\BEA}{\begin{eqnarray}}
\newcommand{\EEA}{\nonumber\end{eqnarray}}

\def\chpt{\raise0.4ex\hbox{$\chi$}PT}
\def\schpt{S\raise0.4ex\hbox{$\chi$}PT}
\def\rschpt{rS\raise0.4ex\hbox{$\chi$}PT}
\def\figref#1{Fig.~\ref{fig:#1}}

\def\secref#1{Sec.~\ref{sec:#1}}

\def\gtwid{{\,\raise.3ex\hbox{$>$\kern-.75em\lower1ex\hbox{$\sim$}}\,}}
\def\ltwid{{\,\raise.3ex\hbox{$<$\kern-.75em\lower1ex\hbox{$\sim$}}\,}}

\def\eqn#1{\label{eq:#1}}

\def\eqs#1#2{Eqs.~(\ref{eq:#1}) and (\ref{eq:#2})}

\def\aschpt{HMrAS\raise0.4ex\hbox{$\chi$}PT}

\title{Charmed and light pseudoscalar meson decay constants from HISQ simulations}

\ShortTitle{\vspace{-2mm} Charmed and light pseudoscalar meson decay constants from HISQ simulations}

\author{\vspace{-3mm}
A.~Bazavov$^a$ \hspace{-1.0mm}\thanks{Present address: Department of Physics and Astronomy, University of Iowa, Iowa City, IA 52240 USA},
C.~Bernard$^b$,
C.~Bouchard$^c$,
C.~DeTar$^d$,
D.~Du$^e$,
A.X.~El-Khadra$^f$,
J.~Foley$^d$,
E.D.~Freeland$^g$,
E.~G\'amiz$^h$,
Steven~Gottlieb$^i$,
U.M.~Heller$^j$,
J.~Kim$^k$ \hspace{-1.0mm}\thanks{Present address: Department of Physics and Astronomy, Seoul National University, Seoul, 151-747, South Korea},
\speaker{J.~Komijani} \nolinebreak $^b$,
A.S.~Kronfeld$^l$,
J.~Laiho$^e$,
L.~Levkova$^d$,
P.B.~Mackenzie$^l$,
E.T.~Neil$^{m,n}$,
J.N.~Simone$^l$,
R.L.~Sugar$^o$,
D.~Toussaint$^k$,
R.S.~Van~de~Water$^l$, and
R.~Zhou$^l$
\\
\llap{$^a$} Physics Department, Brookhaven National Laboratory, Upton, NY 11973, USA\\
\llap{$^b$} Department of Physics, Washington University, St. Louis, MO 63130, USA\\
\llap{$^c$} Department of Physics, The Ohio State University, Columbus, OH 43210, USA\\
\llap{$^d$} Department of Physics and Astronomy, University of Utah, Salt Lake City, UT 84112, USA\\
\llap{$^e$} Department of Physics, Syracuse University, Syracuse, NY 13244, USA\\
\llap{$^f$} Physics Department, University of Illinois, Urbana,  IL 61801, USA\\
\llap{$^g$} Liberal Arts Department, School of the Art Institute of Chicago, Chicago, IL 60603, USA\\
\llap{$^h$} CAFPE and Departamento de F\'isica Te\'orica y del Cosmos, Universidad de Granada, E-18071 Granada, Spain\\
\llap{$^i$} Department of Physics, Indiana University, Bloomington, IN 47405, USA\\
\llap{$^j$} American Physical Society, One Research Road, Ridge, NY 11961, USA\\
\llap{$^k$} Physics Department, University of Arizona, Tucson, AZ 85721, USA\\
\llap{$^l$} Fermi National Accelerator Laboratory\thanks{Operated by Fermi Research Alliance, LLC, 
under Contract No.~DE-AC02-07CH11359 with the US DOE.}, Batavia, IL 60510 USA\\
\llap{$^m$} Department of Physics, University of Colorado, Boulder, CO 80309, USA\\
\llap{$^n$} RIKEN-BNL Research Center, Brookhaven National Laboratory, Upton, NY 11973, USA\\
\llap{$^o$} Department of Physics, University of California, Santa Barbara, CA 93106, USA

\vspace{2mm}
{\large\bf Fermilab Lattice and MILC Collaborations}
\vspace{3mm}

E-mail:
\email{jkomijani@physics.wustl.edu, cb@wustl.edu, doug@physics.arizona.edu}
}

\abstract{
We compute the leptonic decay constants  $f_{D^+}$, $f_{D_s}$, and $f_{K^+}$, and the quark-mass ratios
$m_c/m_s$ and $m_s/m_l$ in unquenched lattice QCD.
We use the MILC highly improved staggered quark (HISQ) ensembles with four dynamical quark flavors. 
Our primary results are
$f_{D^+} = 212.6(0.4)({}^{+1.0}_{-1.2})\ \mathrm{MeV}$,
$f_{D_s} = 249.0(0.3)({}^{+1.1}_{-1.5})\ \mathrm{MeV}$, and
$f_{D_s}/f_{D^+} = 1.1712(10)({}^{+29}_{-32})$,
where the errors are statistical and total systematic, respectively.
We also obtain $f_{K^+}/f_{\pi^+} = 1.1956(10)({}^{+26}_{-18})$, updating our previous result, 
and determine the quark-mass ratios 
$m_s/m_l = 27.35(5)({}^{+10}_{-7})$ and
$m_c/m_s = 11.747(19)({}^{+59}_{-43})$.
When combined with experimental measurements of the decay rates, our results lead to precise determinations of the 
CKM matrix elements $|V_{us}| = 0.22487(51) (29)(20)(5)$, $|V_{cd}|=0.217(1) (5)(1)$ and $|V_{cs}|= 1.010(5)(18)(6)$, 
where the errors are from this calculation of the decay constants, the uncertainty in the experimental decay rates, 
structure-dependent electromagnetic corrections, and, in the case of $|V_{us}|$, the uncertainty in $|V_{ud}|$, respectively.
}

\FullConference{The 32nd International Symposium on Lattice Field Theory\\
                 23-28 June, 2014\\
                 Columbia University New York, NY}

\begin{document}

\section{Introduction}
\vspace{-2mm}
The leptonic decays  of pseudoscalar mesons enable precise determinations of the CKM quark-mixing matrix elements within the Standard Model.
In  particular, experimental rates for the decays $D^+\to\mu^+\nu$, $D_s\to\mu^+\nu$ and $D_s\to\tau^+\nu$, 
when combined with lattice calculations of the charm-meson decay constants  $f_{D^+}$  and $f_{D_s}$,
allow one to obtain $|V_{cd}|$ and $|V_{cs}|$.  
Indeed, this approach results in the most precise current determination of $|V_{cd}|$ to date. 
Similarly, the light-meson decay-constant ratio $f_{K^+}/f_{\pi^+}$ can be used to extract $|V_{us}|/|V_{ud}|$ from 
the experimental ratio of kaon and pion leptonic decay widths \cite{Marciano:2004uf,FPI04}.

We use the lattice ensembles generated by the MILC Collaboration with four flavors ($n_f=2+1+1$) of dynamical quarks using 
the highly improved staggered quark (HISQ) action, and a one-loop tadpole improved Symanzik improved gauge
action~\cite{HPQCD_HISQ,milc_hisq,scaling09,HISQ_CONFIGS}.  
Our data set includes ensembles with four values of the lattice spacing ranging from approximately 0.15~fm to 0.06~fm, 
enabling good control over the continuum extrapolation.
The data set includes both ensembles with the light (up-down), strange, and charm sea-masses close to their physical values
(``physical-mass ensembles'') and ensembles where either the light sea-mass is heavier than in nature, or the
strange sea-mass is lighter than in nature, or both.  

Preliminary results for the charm decay constants and quark masses were presented in Ref.~\cite{LATTICE13_FD}, 
and the final results are given in Ref.~\cite{HISQ_fDfDs_2014}.
These proceedings summarize the analysis and results of Ref.~\cite{HISQ_fDfDs_2014}.
For details about the lattice ensembles used in our calculation and the method for 
extracting the decay constants from two-point correlation functions see Ref.~\cite{HISQ_fDfDs_2014}.

\vspace{-2mm}
\section{Determination of decay constants and quark-mass ratios}
\label{sec:analysis}
\vspace{-2mm}
This section describes the details of the analyses that produce our results for light-light and 
heavy-light decay constants and the ratios of quark masses.  We perform two versions of the analysis. 
The first, the ``physical-mass analysis'' described in \secref{physical-mass-analysis}, is a straightforward procedure 
that essentially uses only the physical-quark mass ensembles.  
On these ensembles, a chiral extrapolation is not needed: 
only interpolations are required in order to find the physical quark-mass point.  
The physical-mass analysis produces our results for quark-mass ratios and $f_{K^+}/f_{\pi^+}$, as well as some additional
intermediate quantities required for the chiral analysis of the $D$ meson decay constants.
The second analysis of charm decay constants, described in \secref{chiral-analysis},  
uses chiral perturbation theory to perform a combined fit to all of our physical-mass and unphysical-mass data, 
and to thereby significantly reduce the statistical uncertainties of the results, especially for $f_{D^+}$.
We take the statistically more precise values of $f_{D^+}$, $f_{D_s}$, and their ratio from the chiral analysis as our final results, 
and use the differences from the results of the simpler physical-mass analysis for estimates of systematic errors.

\vspace{-2mm}
\subsection{Simple analysis  from physical quark-mass ensembles}
\label{sec:physical-mass-analysis}
\vspace{-2mm}
In the first stage of the analysis, we determine tuned quark masses and the lattice spacing (using $f_{\pi^+}$ to fix the scale) 
for each ensemble, and then find the decay constants by interpolation or extrapolation in valence-quark mass to these corrected quark masses.  
Since the decay amplitude $F$ depends on the valence-quark mass, and we wish to use $f_{\pi^+}=130.41$ MeV \cite{PDG} to set the lattice scale, 
we must determine the lattice spacing and tuned light-quark  mass simultaneously.
To do so, we find the light valence-quark mass where the mass and amplitude of the pseudoscalar meson with degenerate valence quarks 
have the physical ratio of $M_\pi^2/f_{\pi^+}^2$.  
(Actually we adjust this ratio for finite size effects. See  Ref.~\cite{HISQ_fDfDs_2014} for more details.)
With the tuned light-quark mass determined, we use the decay amplitude at this mass, $f_{\pi^+}$, to fix the lattice spacing.
In performing the interpolation or extrapolation of $M_\pi^2/f_\pi^2$ we use points with degenerate light valence-quark 
mass $m_{\rm v}$ and employ a continuum, partially quenched, SU(2) \chpt\ form \cite{Sharpe:1997by,Aubin:2003mg}.

We then fix the tuned strange quark mass to the mass that gives the correct $2M_K^2-M_\pi^2$. 
Next we determine the up-down quark mass difference, and hence the up and down quark masses using the difference in $K^0$ and $K^+$ masses.
In this stage of the tuning, the kaon mass is corrected for finite volume effects, electromagnetic effects and isospin breaking effects.
The tuned charm quark mass is also determined from the experimental value of $M_{D_s}$. 

Now that we have found the lattice spacing and tuned quark masses, we can find decay constants 
and masses of other mesons by interpolating or extrapolating to these quark masses.  
We determine the useful quantity $F_{p4s}$ \cite{HISQ_CONFIGS}, which is the light-light pseudoscalar decay constant $F$ evaluated 
at a fiducial point with both valence masses equal to $m_{p4s}\equiv0.4m_s$ and physical sea-quark masses.  
The meson mass at the same fiducial point, $M_{p4s}$, as well as the ratio $R_{p4s}\equiv F_{p4s}/M_{p4s}$, are similarly determined. 
The unphysical decay constant $F_{p4s}$ provides an extremely precise and convenient quantity to set the relative scale
in the chiral analysis, while we use $R_{p4s}$ to tune the strange sea-quark mass.

In the second stage, we combine the results from the individual ensembles and fit to a function of the lattice spacing to find the continuum limit.  
We use the ensembles with unphysical sea-quark masses to make small adjustments for the fact that the sea-quark masses are slightly mistuned. 

Fitting to the lattice spacing dependence is straightforward, because the results from each ensemble are statistically independent.
We perform continuum extrapolations for the ratios of quark masses, $m_u/m_d$, $m_s/m_l$, and $m_c/m_s$,  
for the ratios of decay constants $F_{p4s}/f_{\pi^+}$, $f_{K^+}/f_{\pi^+}$, $f_{D^+}/f_{\pi^+}$,  $f_{D_s}/f_{\pi^+}$, and $f_{D_s}/f_{D^+}$,
and for $M_{p4s}$ and $R_{p4s}$. 
The extrapolated value for $f_{K^+}/f_{\pi^+}$ is our result for this quantity.
Statistical errors on these quantities are estimated with a jackknife method.
We have six continuum extrapolations for each quantity,
which are used to estimate their systematic errors, and to inform the systematic error of the chiral analysis.
The values for the charm-meson decay constants provide consistency checks
on the analysis in \secref{chiral-analysis}, and the spread in
continuum values among the different extrapolations is included in
our estimates of the systematic uncertainty from the continuum extrapolation.

\vspace{-2mm}
\subsection{Chiral perturbation theory analysis of $f_D$ and $f_{D_s}$
} \label{sec:chiral-analysis}
\vspace{-2mm}
Relative scale setting in the combined chiral analysis is done using $F_{p4s}$.  
We use a mass-independent scale-setting scheme. We first determine $aF_{p4s}$ and the quark mass $am_{p4s}$ on the physical-mass ensembles; 
then, by definition, all ensembles at the same $\beta$ as a given physical-mass ensemble have a lattice spacing $a$ and 
value of $am_{p4s}$ equal to those of the physical-mass ensemble. 
To determine $aF_{p4s}$ and $am_{p4s}$ accurately, data is adjusted for mistunings in the sea-quark masses \cite{HISQ_fDfDs_2014}.

The formulas used for the chiral fits and describing our method for
incorporating discretization effects into the extrapolation are discussed in Refs.~\cite{HISQ_fDfDs_2014} and \cite{Bernard-Komijani}.
Our chiral expansion is systematic through NLO.
However, because the high degree of improvement in the HISQ action drastically reduces the coefficient of the leading discretization errors, 
higher order errors are also apparent.
Therefore, we need to consider several parameters related to the discretization effects, which are formally NNLO.
We get acceptable fits when some, but not all, of these parameters are dropped, especially if the coarsest ensembles are omitted.

We have a total of 18 acceptable ($p\geq0.1$) versions of the continuum/chiral fits, which keep or drop the coarsest ensembles,
keep or drop some of the higher order discretization terms, 
and constrain higher order chiral terms and/or discretization terms with priors or leave them unconstrained.
(For the complete list see Ref.~\cite{HISQ_fDfDs_2014}.)
We also have the six versions of the continuum extrapolations used in the tuning procedure that leads to the inputs of quark mass and lattice scale.
This gives a total of 108 versions of the analysis.
We then choose our ``central fit'' simply by requiring that it be a fit to all ensembles and that 
it give results for $\Phi_{D+}$ and $\Phi_{D_s}$ that are as close as possible to the center of the histograms for these quantities.

The central fit provides us with the central values of all output quantities. 
To determine the total statistical error of each output quantity, we use a jackknife method, 
dropping some configurations in turn from each ensemble, and recomputing the inputs (from the physical-mass analysis) as well as the chiral fits.
One can use the total of 108 versions of the analysis to determine the systematic error associated with the continuum extrapolation
(and chiral interpolation) of the charm decay constants in the chiral perturbation theory analysis.

\vspace{-2mm}
\section{Results and conclusions}
\label{sec:conclusions}
\begin{figure}[t]
\begin{center}
        \null\vspace{-08mm}
        \begin{tabular}{l l}
        \null\hspace{-10mm}\includegraphics[trim=0.5in 0 0.3in 0, clip, width=8.3cm]{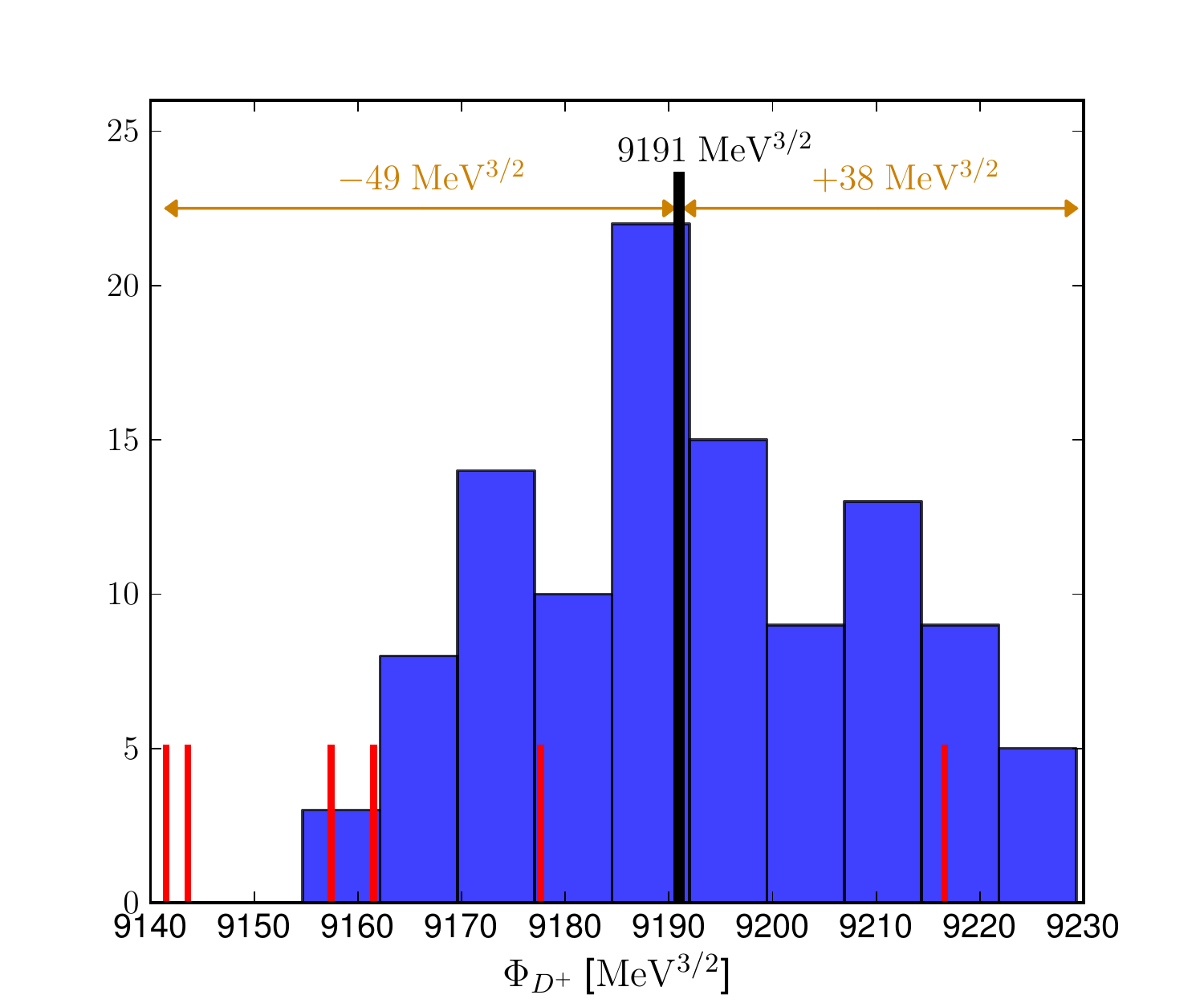}
        &\hspace{-2mm}
        \includegraphics[trim=0.5in 0 0.3in 0, clip, width=8.3cm]{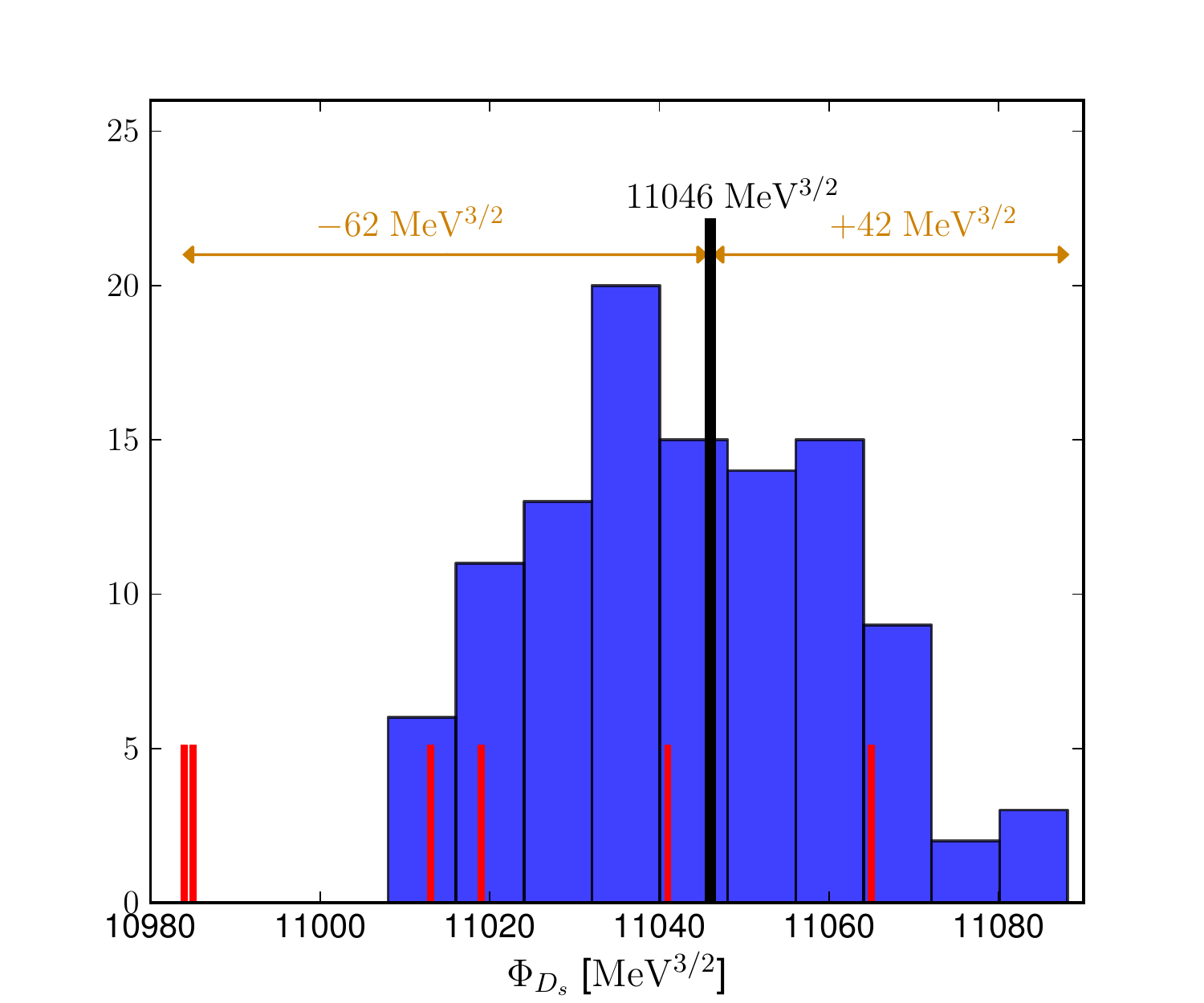}
        \end{tabular}
        \end{center}\vspace{-08mm}
\caption{
Histograms of $\Phi_{D^+}$ and $\Phi_{D_s}$ from the chiral analysis have been overlaid 
with results from various continuum extrapolations in the physical-mass analysis, shown as vertical red lines. 
We take the full ranges shown at the top of each plot as the final estimates of the
systematic errors coming from the continuum extrapolation.
\label{fig:hist_phi_overlay}}
\end{figure}

\vspace{-2mm}
Our main results are for the charm decay constants and their ratio.  We take the results of the central chiral fit
for our best estimate of the central values and statistical errors.
We then use the results of both the physical-mass analysis and the chiral analysis to help
estimate the systematic uncertainties.  
Conservatively, we take the maximum difference seen in these results,
shown in \figref{hist_phi_overlay},
with our central values as the estimate of the continuum extrapolation errors. 
 With this procedure for estimating the systematic uncertainties, 
 the principal role of the combined analysis of physical and unphysical data is to reduce the statistical error.

The analysis on the physical-mass ensembles also helps us estimate the finite-volume and EM errors.
This procedure yields our final results for $f_{D^+}$, $f_{D_s}$ and $f_{D_s}/f_{D^+}$:
\begin{eqnarray}
f_{D^+} &=& 212.6 \pm0.4_{\rm stat}\;{}^{+0.9}_{-1.1}\vert_{a^2\,{\rm extrap}}\pm 0.3_{\rm FV}
\pm 0.1_{\rm EM}\pm 0.3_{f_\pi\, {\rm PDG}}    \ {\rm MeV} ,\eqn{fD-result}\\ 
f_{D_s} &=& 249.0\pm 0.3_{\rm stat}\;{}^{+1.0}_{-1.4}\vert_{a^2\,{\rm extrap}}\pm 0.2_{\rm FV}
\pm0.1_{\rm EM} \pm0.4_{f_\pi\, {\rm PDG}}
\  {\rm MeV} , \eqn{fDs-result}\\ 
f_{D_s}/f_{D^+} &=& 1.1712(10)_{\rm stat}({}^{+28}_{-31})_{a^2\,{\rm extrap}}(3)_{\rm FV}
(6)_{\rm EM} \,.  \eqn{ratio-result} \end{eqnarray}
For the effects of isospin violation we find
\begin{equation} 
f_{D^+}- f_D =
0.47 (1)_{\rm stat} ({}^{+25}_{-\phantom{0}4})_{a^2\,{\rm extrap}}(0)_{\rm FV}  (4)_{\rm EM}  \ {\rm MeV} , 
\eqn{fDp-fD-result}
\end{equation}
where $f_{D}$ is the value of $f_{D^+}$ in the isospin limit, when the light valence mass is equal to $m_l = (m_u+m_d)/2$ instead of $m_d$.
Our result for $f_{D_s}$ is more precise than previous determinations primarily for two reasons.  
First, the statistical errors in our data points for the decay amplitudes are two or more
times smaller than those obtained by others. 
Second, our use of ensembles with the physical light-quark mass eliminates the significant (although not dominant) uncertainty
from  the chiral extrapolation. For $f_{D^+}$ and $f_{D_s}/f_{D^+}$, we also have significantly smaller
continuum-extrapolation errors due to the use of the HISQ sea-quark action and lattice spacings down
to $a \approx 0.06$~fm. 
Moreover, the statistical error in $f_{D^+} $ is slightly more than a factor of
two smaller with the chiral analysis than in the physical-mass analysis.
In fact, the main effect of the chiral analysis on the final results is a significant reduction of the statistical errors,
in particular for $f_{D^+}$.

We also update our determination of the decay-constant ratio $f_{K^+}/f_{\pi^+}$ in Ref.~\cite{FKPRL} from the
physical-mass analysis,
and include results for quark-mass ratios coming from the tuning
procedure and continuum extrapolation described in Sec.~\ref{sec:physical-mass-analysis}:
\begin{eqnarray}
f_{K^+}/f_{\pi^+} &=& 1.1956 (10)_{\rm stat}\;{}^{+23}_{-14}\vert_{a^2\,{\rm extrap}} (10)_{\rm FV} (5)_{\rm
EM}\eqn{fkpiratio-result} \\
m_s/m_l &=& 27.352 (51)_{\rm stat}\;{}^{+80}_{-20}\vert_{a^2\,{\rm extrap}} (39)_{\rm FV} (55)_{\rm
EM}\eqn{slratio-result}\\ 
m_c/m_s &=& 11.747 (19)_{\rm stat}\;{}^{+52}_{-32}\vert_{a^2\,{\rm extrap}}
(6)_{\rm FV} (28)_{\rm EM}\eqn{csratio-result} \,.  \end{eqnarray}
Although our analysis also determines $m_u/m_d$, we do not quote a final result, because the errors in this
ratio are dominated by electromagnetic effects. For more details see Refs.~\cite{HISQ_fDfDs_2014} and \cite{CB-Lat14}.

\vspace{-2mm}
\section{Impact on CKM phenomenology} \label{sec:CKM}
\vspace{-2mm}
We now use our decay constant results to obtain values for CKM matrix elements within the Standard Model,
and to test the unitarity of the first and second rows of the CKM matrix. 
The decay-constant ratio $f_{K^+}/f_{\pi^+}$ can be combined with experimental measurements of the
corresponding leptonic decay widths to obtain a precise value for the ratio
$|V_{us}|/|V_{ud}|$~\cite{Marciano:2004uf}. Combining our updated result for $f_{K^+}/f_{\pi^+}$ 
with recent experimental results for the leptonic branching
fractions~\cite{PDG} and an estimate of the hadronic structure-dependent EM correction~\cite{Antonelli:2010yf},  we obtain 
\begin{equation} |V_{us}|/|V_{ud}| = 0.23081 (52) _{\rm LQCD} (29) _{{\rm BR}(K_{\ell 2})} (21)_{\rm EM} \,. 
\end{equation}
Taking $|V_{ud}|$ from nuclear $\beta$ decay~\cite{Hardy:2008gy}, we also obtain
\begin{equation} |V_{us}| = 0.22487 (51) _{\rm LQCD} (29) _{{\rm BR}(K_{\ell 2})} (20)_{\rm EM} (5)_{V_{ud}}\,. 
\end{equation}
This result for $|V_{us}|$ is more precise than our recent determination from a calculation of the kaon
semileptonic form factor on the physical-mass HISQ ensembles~\cite{Bazavov:2013maa}, and larger by1.8$\sigma$.  
We find good agreement with CKM unitarity, and obtain a
value for the sum of squares of elements of the first row of the CKM matrix consistent with the
Standard-Model prediction zero at a level below $10^{-3}$: 
\begin{equation} 1 - |V_{ud}|^2 - |V_{us}|^2 - |V_{ub}|^2 = 0.00026 (51) \,.  \end{equation}
(Note that $|V_{ub}|^2 \approx 10^{-5}$ does not contribute at the current level of precision.)
Thus our result places stringent constraints on new-physics scenarios that would lead to deviations from first-row CKM unitarity.  
Now that the uncertainty in $|V_{us}|^2$ is approximately the same as that in $|V_{ud}|^2$, 
it is especially important to scrutinize the current uncertainty estimate for $|V_{ud}|$. 

For the determinations of $|V_{cd}|$ and $|V_{cs}|$ given here,
as discussed in Ref.~\cite{HISQ_fDfDs_2014}, we first adjust the experimental decay rates
quoted in the PDG by the known long-distance and short-distance electroweak corrections.  
We then add an estimate of the uncertainty due to the unknown hadronic structure-dependent EM corrections.
With these assumptions, and using our results for $f_{D^+}$ and $f_{D_s}$ from \eqs{fD-result}{fDs-result}, we obtain
\begin{eqnarray} |V_{cd}| &=& 0.217 (1) _{\rm LQCD} (5) _{\rm expt}  (1)_{\rm EM} \,,         \label{eq:Vcd} \\
|V_{cs}| &=& 1.010 (5) _{\rm LQCD} (18) _{\rm expt}  (6)_{\rm EM}                              \label{eq:Vcs} \,, \end{eqnarray}
where EM denotes the error due to unknown structure-dependent EM corrections. 
(See Ref.~\cite{HISQ_fDfDs_2014} for more details.)
In both cases, the uncertainty is dominated by the experimental error in the branching fractions. 
Thus the significant improvement in $f_{D^+}$
and $f_{D_s}$ does not, at present, lead to direct improvement in $|V_{cd}|$ and $|V_{cs}|$.  Experimental
measurements of the $D^+$ decay rates have improved recently~\cite{Rosner:2013ica}, however, such that the
error on $|V_{cd}|$ from leptonic $D^+$ decays is now approximately half that of $|V_{cd}|$ obtained from
either neutrinos~\cite{PDG} or semileptonic $D \to \pi \ell \nu$ decay~\cite{Na:2011mc}.   

Our result for $|V_{cd}|$ agrees with the determination from neutrinos. Our $|V_{cd}|$ is 1.0$\sigma$ lower than the determination 
from semileptonic $D$ decay in Ref.~\cite{Na:2011mc}, while our $|V_{cs}|$ is 1.1$\sigma$ higher than that of Ref.~\cite{Na:2010uf}.
We obtain a value for the sum of squares of elements of the second row of the CKM matrix of 
\begin{equation} 1 - |V_{cd}|^2 - |V_{cs}|^2 - |V_{cb}|^2 = -0.07(4) \,, \end{equation}
showing some tension with CKM unitarity.
(Note that $|V_{cb}|^2 \approx 2\times10^{-3}$ does not contribute at the current level of precision.)
This test will continue to become more stringent as
experimental measurements of the $D^+$ and $D_s$ decay rates become more precise.  At present, even if our
rough estimate of the uncertainty due to structure-dependent EM corrections in Eqs.~(\ref{eq:Vcd}) and~(\ref{eq:Vcs})
is too small by a factor of two,  the errors on $|V_{cd}|$ and
$|V_{cs}|$ would not change significantly.  It will be important, however, to obtain a more reliable estimate of the contributions to
charged $D$ decays due to hadronic structure in the future.   

\vspace{-2mm}
\acknowledgments
\vspace{-2mm}
This work was supported by the U.S. Department of Energy and National Science Foundation,
by the URA Visiting Scholars' program (A.E-K.),
and by the MINECO, Junta de Andaluc\'{\i}a, and European Commission.
Computation for this work was done at
the Argonne Leadership Computing Facility (ALCF),
the National Center for Atmospheric Research (UCAR), 
Bluewaters at the National Center for Supercomputing Applications (NCSA),
the National Energy Resources Supercomputing Center (NERSC),
the National Institute for Computational Sciences (NICS),
the Texas Advanced Computing Center (TACC),
and the USQCD facilities at Fermilab,
under grants from the NSF and DOE.


\end{document}